# The Formation and Characteristics of Acrylonitrile/Urea Inclusion Compound

Jun-Ting Zou, Yu-Song Wang, Wen-Min Pang, Lei Shi*, Fei Lu

Hefei National Laboratory for Physical Sciences at Microscale, University of Science and Technology of China, Hefei, Anhui 230026, China

* Corresponding author. Tel.: +86-551-63607924; E-mail addresses: shil@ustc.edu.cn

**Abstract:**

The formation process and composition of the acrylonitrile/urea inclusion compounds (AN/UIC) with different aging times and AN/urea molar feed ratios are studied by differential scanning calorimetry (DSC) and X-ray diffraction (XRD). It is suggested that DSC could be one of the helpful methods to determine the guest/host ratio and the heat of decomposition. Meanwhile, the guest/host ratio and heat of decomposition are obtained, which are 1.17 and 5361.53 J·mol$^{-1}$, respectively. It is suggested AN molecules included in urea canal lattice may be packed flat against each other. It is found that the formation of AN/UIC depends on the aging time. XRD results reveal that once AN molecules enter urea lattice, AN/UIC are formed, which possess the final structure. When AN molecules are sufficient, the length of AN molecular arrays in urea canals increases as aging time prolonging until urea tunnels are saturated by AN.

**Key words:** Acrylonitrile, Urea inclusion compound, Molar ratio, Heat of decomposition, Formation Process

## I. INTRODUCTION

During recent years, the inclusion compounds, which can be formed by various types of organic or inorganic components [1], have received much interest due to the significance in theory and application [2, 3]. These inclusion compounds are conveniently considered to consist of two distinct parts, host and guest substructures [4]. The most prominent example for the host system is urea, which mostly forms hexagonal structure and comprises continuous, parallel tunnels [4]. The guest molecules might be either small organic molecules (such as n-alkanes and their derivatives [5, 6]) or different types of polymers [7-9]. In a structure like urea inclusion compounds (UIC), where the guest molecules are densely packed along the tunnels formed by the host molecules, the molar ratio of the components (guest/host ratio) is definite for specific guest molecules and depends on the length of the guest molecule [10].

A wide range of fundamental physicochemical properties of UIC have been characterized over several years, such as the structural characteristics of the host lattice [11] and the molecular behavior of the guest molecules [12]. In pure and applied chemistry, UIC are of great importance since they possess high selectivity with respect to the guest molecules [13, 14]. Besides, UIC can isolate polymer chains based upon polymer conformations, which makes that it is possible to study the conformational and motional behaviors of a single polymer chain in the solid state [15]. Another very important application of UIC in polymer science is the inclusion polymerization (canal polymerization), as an alternative to the Ziegler-Natta coordination polymerization method, which can be used to obtain highly stereoregular polymers [15, 16].

The polymerization of diene monomers in inclusion compounds was first studied in 1956 by Clasen [17]. In 1960, White [18] made an extensive study of such inclusion polymerization in urea canals under irradiation with a beam from X-ray type resonant transformer, and obtained highly stereoregular polyacrylonitrile (PAN).

PAN is one of the most widely used polymers for textiles and precursors of carbon fibers [19]. The latter are particularly suitable to produce high performance polymer matrix composites characterized by high strength, stiffness, and lightweight [20, 21]. Commercially available PAN, usually synthesized by radical polymerization, has no specific stereoregularity along the chain direction [22]. It has been realized that isotacticity of PAN is likely one of the important factors contributing to reduce defects in carbon fiber [23, 24]. Thus, it is essential to select a highly stereoregular acrylonitrile (AN) polymer for carbon fibers [25-27]. A number of methods have been developed to produce highly stereoregular PAN [28, 29]. Among them, moreover, inclusion polymerization in urea canals shows evident preponderance in preparing PAN with isotacticity >80% mm-triads [30].

Until now, however, many fundamental issues are still unresolved completely [31]. Especially, the detailed structural information of acrylonitrile/urea inclusion compound (AN/UIC) is indefinite. The guest/host ratio and the heat of decomposition of AN/UIC have not been accurately determined, which are related to understanding the structure of AN/UIC and determine the optimum canal polymerization conditions ensuring high-quality stereoregular PAN samples [30, 32]. There are many reasons for this situation. Firstly, AN/UIC are prepared by mixing AN with urea, followed by cooling to a low temperature (-78 $^{o}$C to -55 $^{o}$C)[18] for a long aging time. This formation process is gradual and non-quantitative, so it will take a very long time to the end of the formation process. Only after the formation process is complete, the determined guest/host ratio and heat of decomposition of AN/UIC are valid. Secondly, the AN/UIC are not stable at ambient temperature. It will be decomposed irreversibly when the temperature is higher than -45 $^{o}$C [33]. Consequently many common analytic methods cannot be used for AN/UIC. Additionally, an ideal single crystal of AN/UIC, which can be effectively characterized, has not been obtained yet.

In present study, the formation process and composition of the AN/UIC with different aging times and AN/urea molar feed ratios are studied by differential scanning calorimetry (DSC) and X-ray diffraction (XRD). It is suggested that DSC could be one of the helpful methods to determine the guest/host ratio and the heat of decomposition. Meanwhile, the guest/host ratio and heat of decomposition are obtained. The structural detail of AN/UIC has been discussed.

## II. EXPERIMENTS

Acrylonitrile was vacuum distilled followed by fractional distillation over $CaH_2$ under nitrogen and reserved with molecular sieve (4-A). Urea was purified by recrystallization, and other chemicals were analytical reagents and used without further purification.

The AN/UIC samples were prepared by mixing AN with urea at calculated molar ratios, followed by cooling to a low temperature (at least lower than -50 $^o$C) and keeping for a certain aging time.

DSC curves were measured on a DSC Q2000 thermal analyzer (TA Instruments) at a scanning rate of 10 $^o$C/min from -90 $^o$C to 0 $^o$C. The samples (4 mg to 8 mg) were transferred to TA aluminum pans in liquid nitrogen and then tested. An empty pan was used as reference. Thermograms were analyzed by Universal Analysis 2000 software (Version 4.7A, TA Instruments). XRD patterns were measured on a Rigaku TTR III diffractometer with Cu K$\alpha$ ($\lambda$ = 1.54187 Å) radiation at -120 $^o$C. To avoid the decomposition of AN/UIC, the samples were transferred to sample holders in liquid nitrogen and then tested.

## III. RESULTS AND DISCUSSION

### A. DSC curves of AN/UIC

The DSC curves of prepared AN/UIC with different AN/urea molar feed ratios and aging times at the aging temperature -60 $^o$C and just mixed AN-urea are shown in Fig. 1, from which two peaks at around -80 $^o$C and -30 $^o$C, respectively, can be found. The former is melting peak of AN and the latter one is the endothermic peak [30, 33] which is caused by the decomposition of AN/UIC.

Due to the differences of the sample masses for DSC measurement in each time, the absolute values of AN melting peak areas and AN/UIC decomposition peak areas cannot be directly compared with each other. To investigate the composition of AN/UIC, the relative quantity changes of AN with AN/UIC are employed. When the area ratios of AN melting peak to AN/UIC decomposition peak are plotted, decay curves are obtained (Fig. 2), which indicate that free AN molecules in the systems are gradually reduced and included into the urea lattice forming inclusion compounds with increasing aging time. Accordingly, with sufficient AN molecules, the molar ratio of included AN molecules to urea molecules increases with increasing aging time until urea tunnels are saturated by AN and the molar ratio reaches the maximum and unchanged value, which can be regarded as the guest/host ratio of AN/UIC. Likewise, only when urea tunnels are saturated by AN, the heat of decomposition of AN/UIC determined is valid. It is noticeable that in Fig. 1 there is no AN melting peak for the sample with AN/urea molar feed ratio 1/2 and aging time 90 h, which reveals that all the AN molecules in the systems have been included completely in the urea lattice, and the amount of AN molecules may be so insufficient that the finish of the formation process of AN/UIC is untimely. On the contrary, however, for samples with AN/urea molar feed ratio range from 2/1 to 1.5/1, even if the aging time reaches 210 h, AN melting peak is still existing, which suggests that the amount of AN molecules may be excessive. When the aging time is over 210 h, the melting/decomposition peak area ratios of these samples are almost unchanged (Fig. 2), which indicates that the urea tunnels are completely saturated by AN molecules and the formation process of AN/UIC is essentially finished.

The relationship between the melting/decomposition peak area ratio and the aging time at the aging temperature -78 $^\circ$C for the samples with AN/urea molar feed ratio 2/1 is shown in Fig. 3. Like that in Fig. 2, the melting/decomposition peak area ratio decreases with increasing aging time until urea tunnels are saturated by AN, and keeps stable at about 0.44, which is the same as that at -60 $^\circ$C. However, the formation process of AN/UIC at -78 $^\circ$C takes more than 50 days, which is much longer than that at -60 $^\circ$C. It is obvious that the aging temperature shows an evident influence on the formation of AN/UIC.

**B. Determination of guest/host ratio and heat of decomposition of AN/UIC**

As mentioned above, AN/UIC are in a stable equilibrium state when the aging time is over 210 h for the samples with AN/urea molar feed ratio range from 2/1 to 1.5/1. Thus, heat data obtained by DSC for these samples with an aging time of 400 h could be used to determine the guest/host ratio and the heat of decomposition. First of all, the formation process of AN/UIC could be described as follows,

$g$ AN + $h$ Urea → UIC,

$g + h = 1$ (1)

where $g$ is the proportion of AN to AN/UIC, $h$ is the proportion of urea to AN/UIC. For a sample with excessive AN, in quantity, there is a relationship in composition of the sample as following,

AN (free) + AN (included) = $r$ Urea (2)

where $r$ is the molar feed ratio (AN/urea). From the point of energy, the relationship could be depicted as follows:

$$\frac{\Delta H(-80^\circ C)}{\Delta H_{fus}(AN)} + g \times \frac{\Delta H(-30^\circ C)}{\Delta H_{deco}(IC)} = r \times h \times \frac{\Delta H(-30^\circ C)}{\Delta H_{deco}(IC)} \quad (3)$$

where $\Delta H_{fus}(AN)$ is the enthalpy of the fusion of AN, $\Delta H_{deco}(IC)$ is the heat of the decomposition of AN/UIC, $\Delta H(-80^\circ C)$ and $\Delta H(-30^\circ C)$ are heat of AN melting and AN/UIC decomposition respectively. In these variables, $r$ is known. $\Delta H(-80^\circ C)$ and $\Delta H(-30^\circ C)$ can be determined by DSC. $\Delta H_{fus}(AN)$ is 6.23 kJ·mol$^{-1}$ [34]. $g$, $h$ and $\Delta H_{deco}(IC)$ are to be determined. The guest/host molar ratio ($n$) of AN/UIC is equal to $g/h$. Putting the experiment data (listed in Table 1) into formula (3), two dualistic equations could be obtained. Solving the equations by pairs, the final results are obtained as following:

$g$ = 0.54 (±0.02), $h$ = 0.46 (±0.02), and $\Delta H_{deco}(IC)$ = 5361.53 (±200) J·mol$^{-1}$.

Thus, the guest/host molar ratio ($n$) of AN/UIC is obtained, which is 1.17. According to the dimensions of the acrylonitrile molecule (approximately 7.1 Å long, 4.7 Å wide and 3.5 Å thick) [35] and the dimensions of the urea host structure ($a = b \approx 8.2$ Å, $c \approx 11.0$ Å, tunnel diameter about 5.5 Å to 5.8 Å)[36], if AN molecules are packed end to end in the tunnels, $n$ would be 0.3876. The observed guest/host molar ratio is so large that it requires the AN molecules to be packed flat against each other.

There are four topological models (Model I, II, III, IV) have been developed for the estimation of guest/host ratio [37] and three topological (Model V, VI, VII) models for the estimation of heat of decomposition [38] of urea inclusion compounds, in which three commonly used topological descriptors, *molecular connectivity index* ($\chi$), *Wiener's index* ($W$) and *eccentric connectivity index* ($\xi^c$), are employed. These topological descriptors of AN are calculated by VCCLAB [39, 40] ($\chi$ = 1.91, $W$ = 10 and $\xi^c$ = 14). The guest/host molar ratio and $\Delta H_{deco}$ of AN/UIC predicted using these models are listed in Table 2.

It can be found that the values of $n$ and $\Delta H_{deco}$ obtaineded from DSC differ greatly from those values predicted by the topological models. This is possibly because the topological models are developed from a dataset comprising of scores of reported UIC accommodating guest molecules based on a sufficiently long n-alkane chain [37, 38] and the AN molecule is so small comparing with other guest species of UIC that the interaction between host and guest is very weak. Another reason might be that AN molecules arrange in urea in a quaint manner. Typical guest molecules included in UIC are densely packed end to end along the tunnels [1]. As discussed above, if AN molecules are packed end to end in the tunnels, $n$ would be 0.3876, which is close to the values predicted using topological models. This demonstrates from another point of view that AN molecules might be packed flat against each other.

## C. The structure of AN/UIC characterized by XRD

As mentioned above, the free AN molecules in the reaction system are gradually included in the urea lattice with increasing aging time, forming inclusion compounds. The question arises, what are the structural differences between the samples of the AN/UIC saturated by AN (or the tunnels completely occupied by AN molecules) and the urea lattice partly including AN molecules. It is believed that urea itself has a tetragonal lattice structure, but in the presence of lower organic substances such as AN, a crystalline transformation will happen and the hexagonal lattice of urea, with an AN molecular array in the central position, will appear [30]. It has been also shown that if the guest molecules are removed from the inclusion compound, the tunnels will collapse and the urea will recrystallize in its tetragonal lattice structure, which does not contain any tunnels [1].

The XRD patterns of the samples with different aging times are shown in Fig. 4.

The samples have the same AN/urea molar feed ratio of 2/1. According to the assignments proposed by Yoshii [33], the XRD patterns have been indexed. From the XRD patterns, it can be found that the crystal structures of the samples have little change during the formation process, which is different from that of just mixed sample. The structural differences of the just mixed sample and the aging samples, and the almost unchanged structure of the aging samples, indicate that once AN molecules enter urea lattice ($\leq$ 96 h), AN/UIC are formed, which will possess the final hexagonal lattice structure. Moreover, it can be found in Fig.4 that the peak intensities of urea decrease with increasing aging time, which indicates the content of the urea with tetragonal lattice structure is decreased with increasing aging time. The reduction in urea content implies the increase of AN/UIC content and consequently more AN molecules are included in urea canals. That is, when AN molecules are sufficient, the length of AN molecular arrays in urea canals increases as aging time prolonging until urea tunnels are saturated by AN (Fig. 5).

## IV. CONCLUSION

In this paper, the guest/host ratio and heat of decomposition are obtained, which are 1.17 and 5361.53 J·mol$^{-1}$, respectively. It is suggested AN molecules included in urea canal lattice may be packed flat against each other. It is found that the formation of AN/UIC depends on the aging time and ends after enough aging time. Furthermore, it is found that once AN molecules enter urea lattice, AN/UIC are formed, which possess the final structure. When AN molecules are sufficient, the length of AN molecular arrays in urea canals increases as aging time prolonging until urea tunnels are saturated by AN. These results are helpful in understanding the structure of AN/UIC and determining the optimum canal polymerization conditions ensuring high-quality stereoregular PAN samples.

## V. ACKNOWLEDGMENTS

This work was financially supported by the Fundamental Research Funds for the Central Universities (No. WK2340000016), and USTC Innovation Funds for

**TABLE 1.** Data Used in Determination of Guest/Host Ratio and Heat of Decomposition of AN/UIC

| Mole feed ratio (AN/urea), $r$ | $\Delta H(-80^\circ C)$/mJ | $\Delta H(-30^\circ C)$/mJ |
|---|---|---|
| 1.5/1 | 81.87 | 464.1 |
| 2/1 | 239.8 | 539.6 |

**TABLE 2.** Guest/host Ratio and Heat of Decomposition of AN/UIC Predicted Using Topological Models

| Topological models | Topological descriptors | Predicted $n$ | Predicted $\Delta_{deco}H$ ( kJ·mol$^{-1}$) |
|:---:|:---:|:---:|:---:|
| I | $\chi$ | 0.3185 | / |
| II | $W$ | 0.1531 | / |
| III | $\xi^c$ | 0.1835 | / |
| IV | $\chi$ and $W$ | 0.3019 | / |
| V | $\chi$ | / | 13.0494 |
| VI | $W$ | / | 39.4191 |
| VII | $\xi^c$ | / | 31.08 |

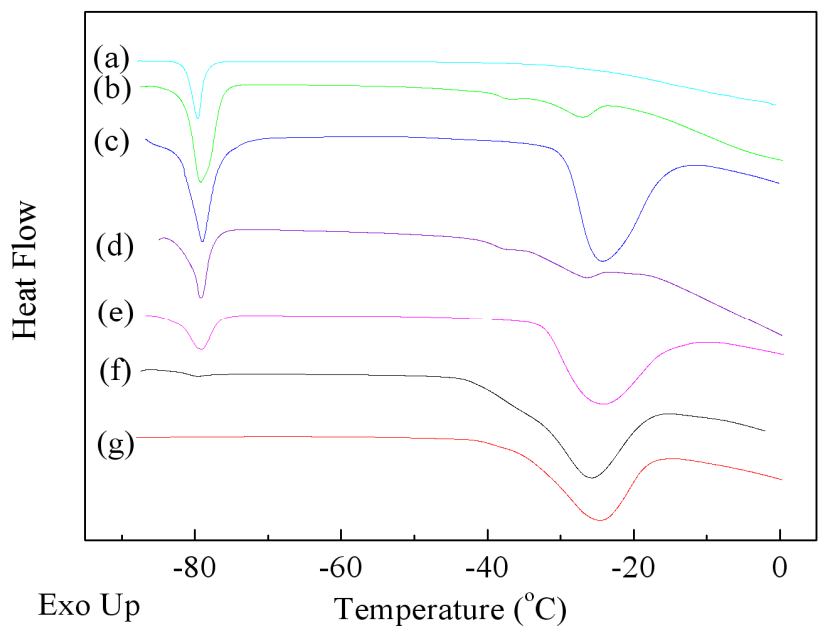

FIG 1. Some typical DSC curves of prepared AN/UIC and just mixed AN-urea: (a), just mixed AN-urea; (b), AN/urea=2/1, aging time 42 h; (c), AN/urea=2/1, aging time 210 h; (d), AN/urea=1.5/1, aging time 42 h; (e), AN/urea=1.5/1, aging time 210 h; (f), AN/urea=1/2, aging time 42 h; (g), AN/urea=1/2, aging time 90 h.

.

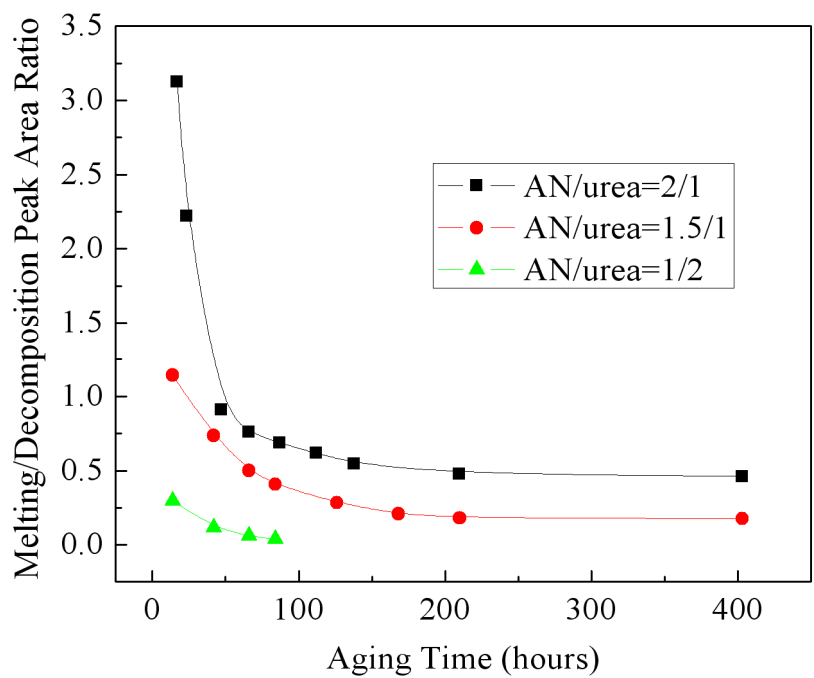

FIG 2. Relationship between the melting/decomposition peak area ratio and the aging time at the aging temperature -60 $^{o}$C.

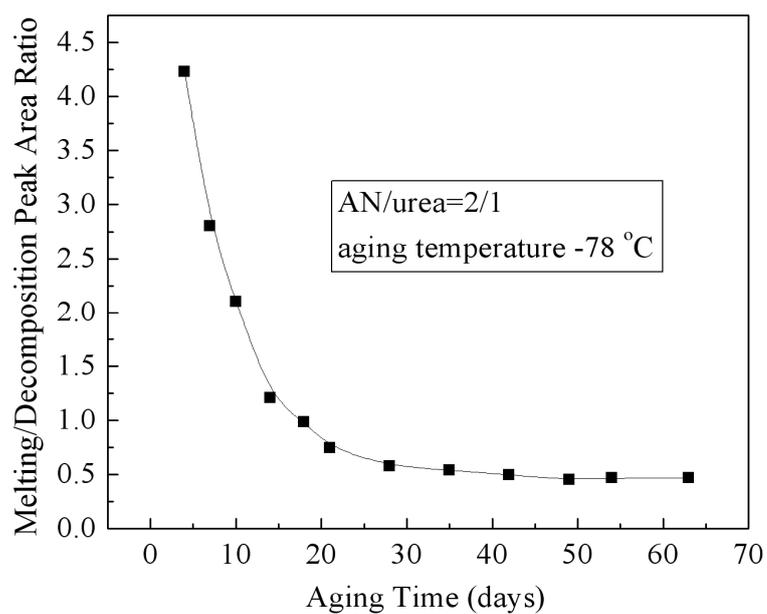

FIG 3 Relationship between the melting/decomposition peak area ratio and the aging time at the aging temperature -78 °C.

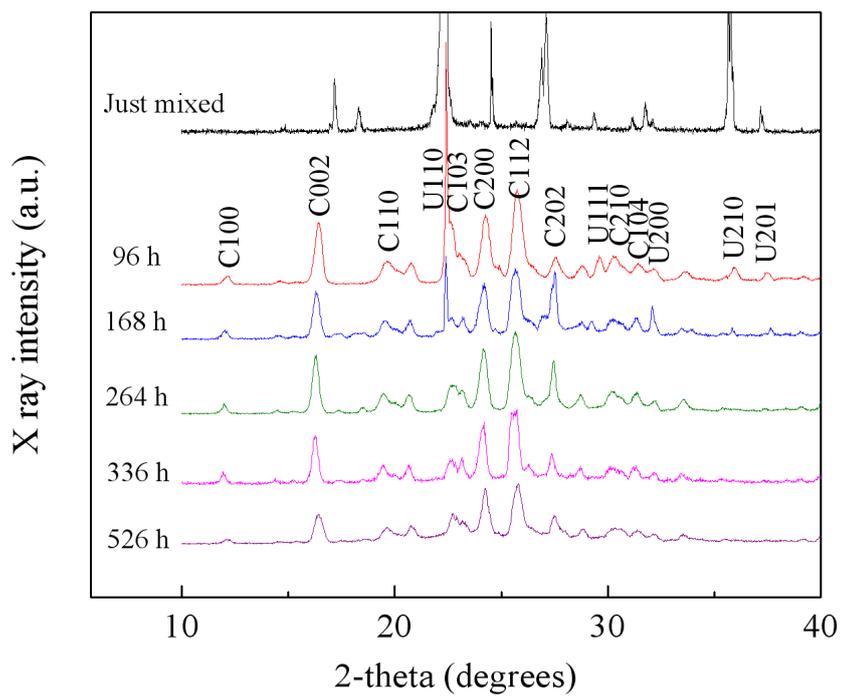

FIG 4. XRD results of samples with different aging times. These samples have the same original AN/urea molar feed ratio 2/1. The notation given in Ref. [33] was used. C: AN/UIC; U: urea.

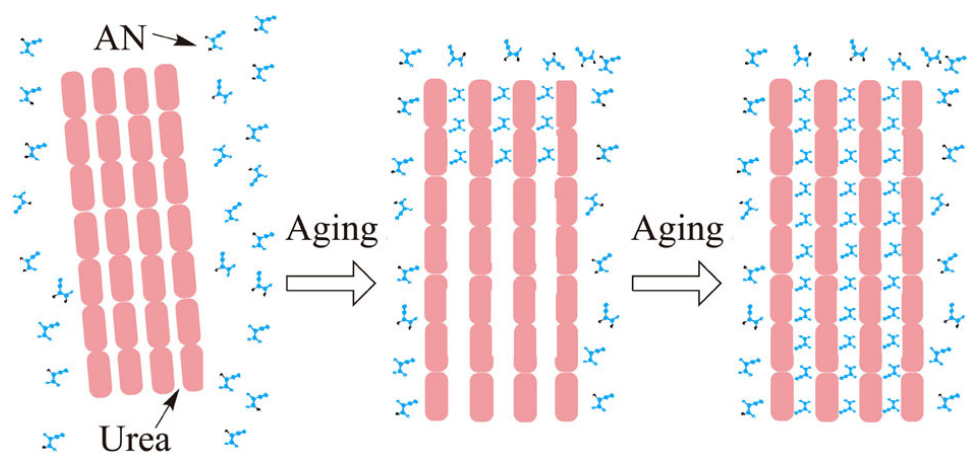
FIG 5. Schematic representation of the formation process of AN/UIC.